\documentclass{article}
\usepackage{arxiv}

\usepackage[automake,toc,abbreviations]{glossaries-extra} 
\usepackage{graphicx} 
\usepackage{amsfonts}
\usepackage{amsmath}
\usepackage{subcaption}
\usepackage[style=ieee]{biblatex}
\usepackage{xcolor}
\usepackage{authblk}
\usepackage[T1]{fontenc}
\usepackage{hyperref}

\DeclareMathOperator*{\argmin}{arg\,min}

\makeglossaries

\title{BLISSNet: Deep Operator Learning for Fast and \\ Accurate Flow Reconstruction from Sparse Sensor Measurements}
\author[1]{Maksym Veremchuk}
\author[1]{K. Andrea Scott}
\author[1]{Zhao Pan}

\affil[1]{Department of Mechanical and Mechatronics Engineering, University of Waterloo} 
\affil[ ]{\texttt{\{mrveremc, ka3scott, zhao.pan\}@uwaterloo.ca}}
\date{}

\begin{document}

\maketitle
\begin{abstract}
Reconstructing fluid flows from sparse sensor measurements is a fundamental challenge in science and engineering. Widely separated measurements and complex, multiscale dynamics make accurate recovery of fine-scale structures difficult. In addition, existing methods face a persistent tradeoff: high-accuracy models are often computationally expensive, whereas faster approaches typically compromise fidelity.
In this work, we introduce BLISSNet, a model that strikes a strong balance between reconstruction accuracy and computational efficiency for both flow reconstruction and nudging-based data assimilation.
The model follows a DeepONet-like architecture, enabling zero-shot inference on domains of arbitrary size. After the first model call on a given domain, certain network components can be precomputed, leading to low inference cost for subsequent evaluations on large domains. Consequently, the model can achieve faster inference than classical interpolation methods such as radial basis function or bicubic interpolation.
This combination of high accuracy, low cost, and zero-shot generalization makes BLISSNet well-suited for large-scale real-time flow reconstruction and data assimilation tasks.
\end{abstract}

\section{Introduction}

In many real-world scenarios, acquiring measurements over an entire large domain is impractical or even impossible~\cite{allsensorsinfeasable}. As a result, directly reconstructing full fields from sparse measurements is a fundamental problem in many scientific and engineering applications. In climate science, accurate interpolation is essential for estimating temperature, precipitation, and other geophysical variables over large regions~\cite{climate_interpolation, climate_interpolation_2, climate_interpolation_3}. Similar challenges arise in flow dynamics, where velocity or pressure fields must be reconstructed from limited sensor data~\cite{flow_interpolation, flow_interpolation_2}. In remote sensing, sparse satellite observations are interpolated to form complete spatial maps~\cite{remote_sensing_interpolation}, while in medical applications, reliable reconstruction from limited measurements is required to reduce acquisition time (e.g., MRI, CT, blood flow)~\cite{medicine_interpolation, medicine_interpolation_2}. Consequently, developing methods that can both quickly and accurately reconstruct continuous fields from incomplete and noisy measurements remains a significant challenge. 

There are two main types of methods for direct flow reconstruction: data-driven methods and classical interpolation techniques. Data-driven methods often achieve high reconstruction accuracy through reduced-order modeling. 
However, their long inference time makes them unsuitable for time-sensitive applications. 
For example, real-time inference is critical in scenarios such as disaster response, including wildfire spread prediction, earthquake monitoring, and radiation surveillance~\cite{rt_inverse_distance_weighting_interpolation}, as well as in ocean state forecasting~\cite{rt_ocean_interpolation} and navigation systems~\cite{rt_navigation_interpolation, rt_navigation_interpolation2}. 
In contrast, classical interpolation methods, such as bicubic interpolation and radial basis function (RBF) interpolation, are computationally efficient and well-suited for real-time use. However, they often lack the accuracy needed for high-fidelity reconstruction in complex physical fields. 

In addition to direct field reconstruction, data assimilation is widely used across many scientific disciplines.
Data assimilation methods typically utilize statistical and physical models to fuse measurements and predictions over time, producing an optimal estimate of the evolving field that observations alone cannot achieve.
This approach is widely used in applications such as weather and ocean forecasting, where sparse observations are combined with model-based estimation to improve both spatial field reconstruction and temporal prediction of system evolution.~\cite {dataassimilation_interpolation, flow_interpolation}. 

In direct flow reconstruction and data assimilation, an inherent accuracy–speed tradeoff arises: increasing reconstruction fidelity generally entails higher computational cost, whereas faster methods often reduce solution quality. 
To address this limitation, we propose a two-stage reconstruction architecture that integrates deep learning with a computational structure suitable for large-scale, real-time deployment. 
The reconstruction is decomposed into components that are precomputed offline and lightweight operators evaluated during inference, thereby transferring the dominant computational burden away from runtime while preserving expressive capacity. 
Within this framework, we introduce BLISSNet, an interpolation operator designed to achieve both high accuracy and computational efficiency. 
On large domains, BLISSNet attains faster inference than classical interpolation schemes while maintaining accuracy comparable to state-of-the-art data-driven approaches. 
The model further supports zero-shot super-resolution, enabling generalization to resolutions not encountered during training. 
These characteristics define a new operating regime for sparse field reconstruction, demonstrating that high accuracy and real-time performance can be achieved simultaneously within a scalable framework suitable for scientific and engineering applications.

\section{Problem statement and SOTA}

Consider $\mathcal{U}$ as a Banach space composed of vector-valued functions, described as follows:
\begin{equation}
    \mathcal{U} = \{u:\mathcal{X} \rightarrow \mathbb{R}^{d_u}\}, \mathcal{X} \in \mathbb{R}^{d_x} 
\end{equation}
where $u$ is an input function.

In our task, we want to recover the function $u$ from the set of sparse observations $\mathbf{Obs}=\{x_i, u_i(x_i)\}^N$, where $x_i \in \mathcal{X}$, $N$ is the number of available observations (sensors), which is typically 1-5\% of the total number of available spatial locations. 

The main objective is to create an interpolation neural operator $\mathcal{\hat{G}}(\cdot, {\theta})$ to approximate a ground-truth interpolator $\mathcal{G}(\cdot)$ that maps the observations to the full field: 
\begin{equation}
    \argmin_{\theta} \mathbb{E} || \mathcal{G}(\mathbf{Obs})-\mathcal{\hat{G}}(\mathbf{Obs}, {\theta})||_{\mathcal{U}}^2,
\end{equation}
where $\theta$ is the parameter set of the neural operator.



A common approach for sparse-to-field interpolation is a DeepONet-style~\cite{deeponet} architecture with a branch and a trunk network. The trunk produces a set of basis functions on the query grid, and the branch maps the observed sensor values to the corresponding coefficients, yielding the reconstruction
\begin{equation}
\mathcal{G}(u_i(x_i))(x) \approx \sum_{k}^{K} \text{branch}(u_i(x_i), \theta)_k \cdot \text{trunk}(x_i, \theta)_k
\end{equation}
where $K$ is the number of basis functions, $\text{branch}$ and $\text{trunk}$ are neural networks with parameters $\theta$.
This factorization lets models trained on one grid evaluate on larger domains by reusing the learned bases. However, DeepONet assumes fixed sensor locations and a fixed input size, so it cannot handle randomly placed sensors or variable sensor counts, which limits its use when sensor layouts change.

Several neural operator architectures have been developed to address variable input size in sparse sensor settings. Variable-Input Deep Operator Networks (VIDON)~\cite{vidon} extend DeepONet by replacing the branch network with a linear attention mechanism that processes the coordinates and values of observed points. Resolution Independent Neural Operator (RINO)~\cite{rino} learns a data-driven basis, projects the input onto this basis to obtain coefficients, transforms these coefficients through a neural network, and reconstructs the field from the updated coefficients and learned basis functions. In addition, the Voronoi-based neural network~\cite{voronoi_interpolation} provides a lightweight baseline by constructing a Voronoi diagram from sensor locations, assigning each cell the corresponding measurement, and refining the result with a shallow CNN. While these approaches are efficient and achieve strong accuracy in moderately sparse regimes, their performance degrades when observations become extremely sparse or when the underlying system exhibits complex dynamics.

Diffusion-based methods have been proposed to reconstruct fields from sparse observations. However, their computational cost is high. 
For example,  works such as DiffusionPDE~\cite{diffusionpde} and physics-informed diffusion~\cite{physics_informer_diffusion} report high accuracy, they require large memory and long sampling chains, which makes them hundreds of times slower than transformer or DeepONet-based operators. 
This latency and computational burden preclude deployment in systems requiring rapid response, and diffusion-based methods are therefore excluded as baselines.



Among state-of-the-art (SOTA) models, the transformer-based Senseiver~\cite{senseiver} and OFormer~\cite{oformer} demonstrate high accuracy.
In detail, both models employ a closely aligned transformer-based architecture in which an encoder first aggregates coordinate-value pairs into a latent representation and a decoder then maps that representation back to a dense spatial grid. Specifically, for OFormer, the encoder 
($\text{Enc}$) ingests sampled inputs to produce features $\Phi$, and the decoder $\text{CA}_{oformer}$ uses cross-attention to query those features at target grid locations:
\begin{equation}
\Phi = \text{Enc}(\mathbf{Obs}),
\label{eq:encoder_oformer}
\end{equation}
\begin{equation}
\hat{u}(\mathbf{Dom}) = \text{CA}_{oformer}(\Phi, \mathcal{F}(\mathbf{Dom}))
\label{eq:decoder_oformer}.
\end{equation}
Here, encoder $\text{Enc}$ consists of Galerkin transformer blocks and encodes the coordinates and observed values into a feature vector $\Phi$. Decoder $\text{CA}_{oformer}$ performs cross-attention from every query position in the domain grid $\mathbf{Dom} = \{x_i\}^{D^2}$ to the encoded features. $\mathcal{F}$ is a Fourier feature mapping for positional encoding~\cite{fourier_positional_encoding}. 

Although Senseiver and OFormer share an effective encoder-decoder paradigm, both rely on cross-attention over the full spatial grid, which is the primary runtime bottleneck. Specifically, projecting $\Phi$ to all $D^2$ query locations (Equation~\eqref{eq:decoder_oformer}) scales with the grid resolution $D^2$, leading to substantial computational and memory overhead in computing the interpolated output $\hat{u}(\mathbf{Dom})$. To address this inefficiency, we propose an architecture that matches the accuracy of OFormer while redesigning the cross-attention mechanism to significantly reduce latency and memory consumption.

\section{Methodology}

\begin{figure}[h!]
    \centering
    \begin{subfigure}[b]{0.98\textwidth}
        \centering
        \includegraphics[width=\linewidth]{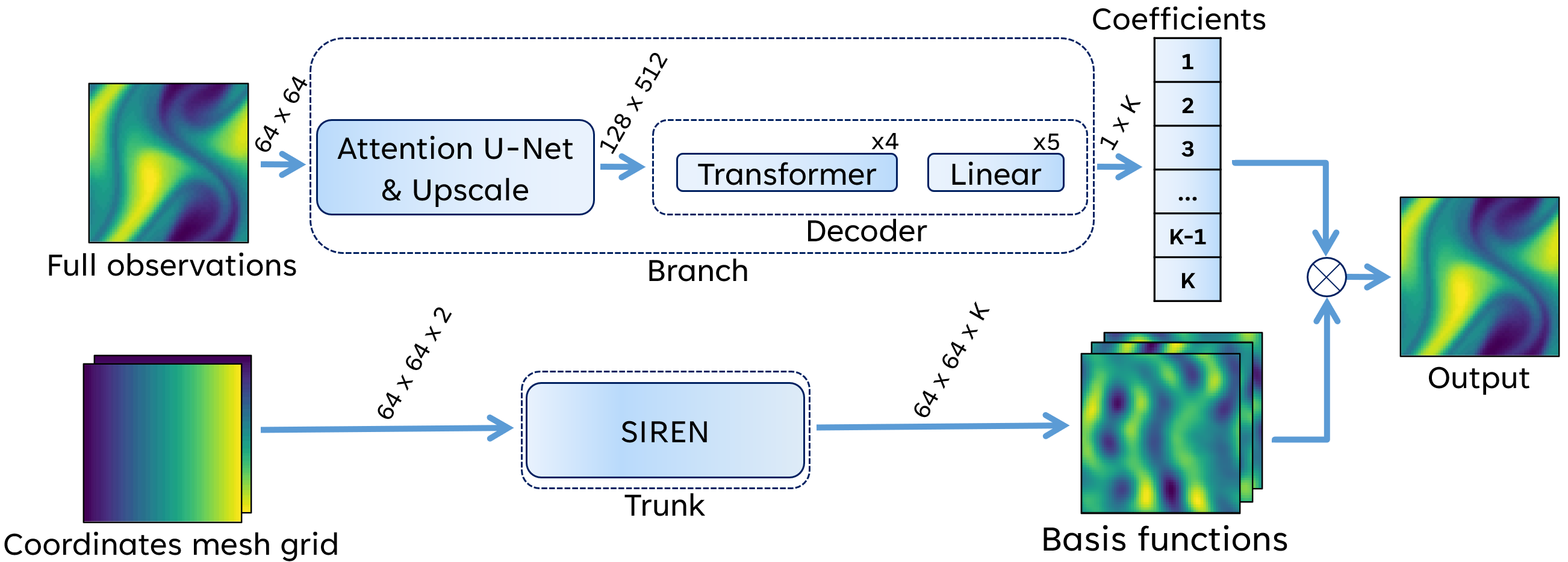}
        \caption{}
        \label{fig:BLISSNet_architechture_stage1}
    \end{subfigure}\hfill\hfill
    \begin{subfigure}[b]{0.98\textwidth} 
        \centering
        \includegraphics[width=\linewidth]{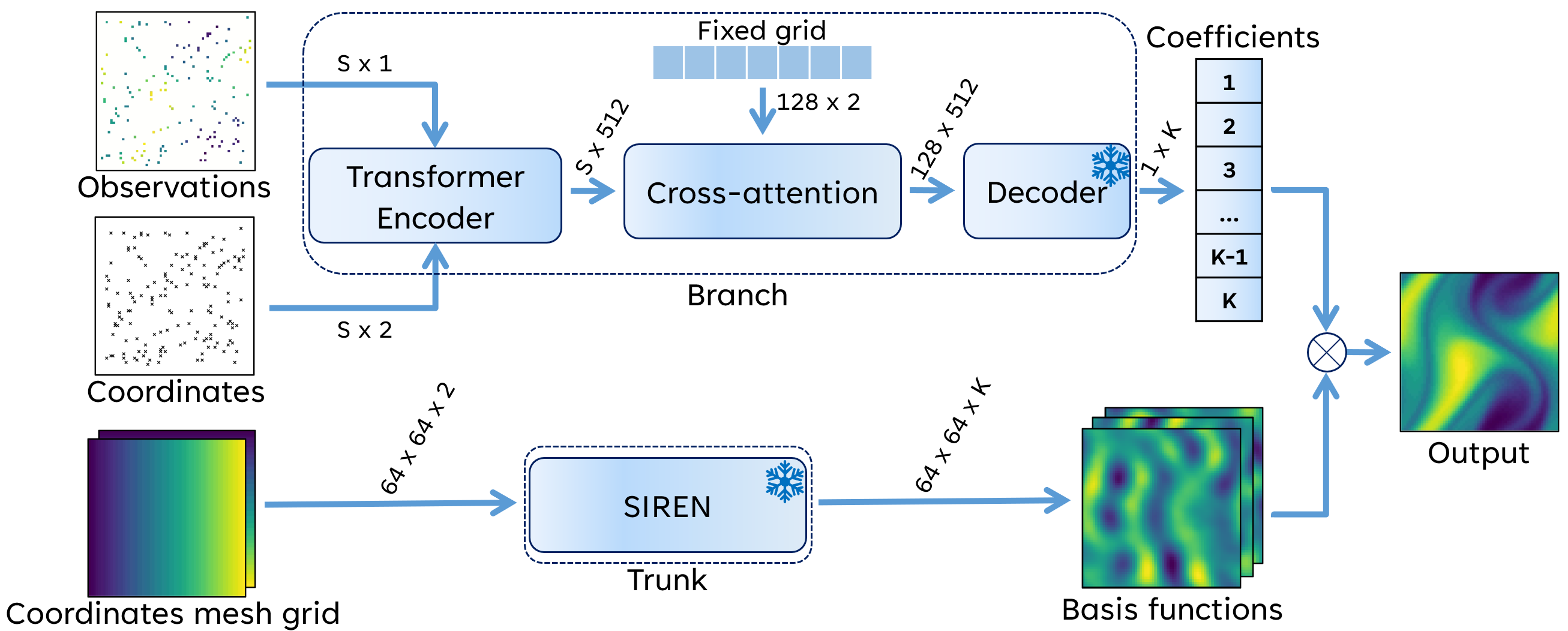}
        \caption{}
        \label{fig:BLISSNet_architechture_stage2}
    \end{subfigure}
    \caption{BLISSNet uses a two-stage architecture. In stage one (panel a), the model is trained on fully observed data to learn a SIREN-based trunk that provides basis functions and a decoder that outputs the coefficients. The reconstruction is a weighted sum of these bases, with $K$ denoting the number of coefficients and basis functions. In stage two (panel b), the model performs interpolation from sparse observations, with $S$ denoting the number of available measurements, and it reuses the SIREN trunk and the decoder learned in stage one while training a Transformer encoder and cross-attention to map sensor coordinates and values to features. We utilize the OFormer encoder (representative of the Transformer encoder) in stage two. However, any suitable encoder can be substituted, and more advanced designs may further enhance interpolation quality. Snowflake here denotes that the weights are frozen during the second stage of training. }
    \label{fig:BLISSNet_architechture}
\end{figure}

\subsection{BLISSNet}\label{sec:blissnet}
To reduce the scaling problem of cross-attention in interpolation, we propose a new model, BLISSNet (\textbf{\underline{B}}asis-functions \textbf{\underline{L}}earned \textbf{\underline{I}}nterpolation from \textbf{\underline{S}}parse \textbf{\underline{S}}ensors \textbf{\underline{Net}}work). BLISSNet is trained using a two-stage approach (Fig.~\ref{fig:BLISSNet_architechture}). In the first stage, the model learns to reconstruct fully observed images, enabling it to acquire strong, dataset-specific representational patterns. In the second stage, the model is trained on partially observed inputs to recover the original images. This stage leverages the representations learned in the first stage, allowing the network to perform more accurate and robust interpolation.

In the first stage (Fig.~\ref{fig:BLISSNet_architechture_stage1}, the trunk network learns a set of basis functions by training on full images where all point values are known. The trunk network is implemented as a SIREN~\cite{siren} model with three linear layers and sine activation functions. Since standard multilayer perceptrons (MLPs) are unable to capture high-frequency information, they are not suitable for the trunk in this case. 

For the branch network in the first stage, we use an Attention U-Net as the encoder to represent the image, followed by a decoder composed of transformer blocks and fully connected layers. This decoder produces a vector of coefficients, which are multiplied by the basis functions learned by the trunk network.

To train the first stage, the loss function is defined as the cumulative mean squared error (MSE) between the ground-truth targets and the reconstructed outputs. The reconstruction is obtained by multiplying the basis functions produced by the trunk network with the corresponding coefficients predicted by the branch network.

In the second stage (Fig.~\ref{fig:BLISSNet_architechture_stage2}, we use the same model architecture as in the first stage, but replace the Attention U-Net encoder with the encoder and cross-attention blocks from OFormer ($\text{Enc}$ and $\text{CA}_{oformer}$ from Equation~\ref{eq:encoder_oformer} and~\ref{eq:decoder_oformer}). The trunk and the decoder at the end of the branch are frozen during this stage, allowing the model to reuse the weights learned in the first stage. This ensures that the basis functions representing the dataset, as well as the decoder for the corresponding coefficients, remain optimal while the encoder learns to map sparsely observed inputs to these pre-trained representations.

To alleviate the computational bottleneck inherent to the cross-attention decoder $\text{CA}_{oformer}$ our design employs a fixed mesh grid in new cross-attention $\text{CA}_{\text{bliss}}$ with additional coefficient decoder $\text{Dec}_\text{coef}$, which yields a constant number of coefficients $K$ instead of generating outputs over the full domain mesh:
\begin{equation}
\Phi = \text{Enc}(\mathbf{Obs}),
\label{eq:encoder_bliss}
\end{equation}
\begin{equation}
 [c_1, c_2 ... c_K] = \text{Dec}_{\text{coef}}(\text{CA}_{\text{bliss}}(\Phi, \mathcal{F}(\text{FG}))),
\label{eq:decoder_bliss}
\end{equation}
\begin{equation}
[f_1(\mathbf{Dom}), f_2(\mathbf{Dom})...f_K(\mathbf{Dom})] = \text{SIREN}(\mathbf{Dom}),
\label{eq:siren_bliss}
\end{equation}
\begin{equation}
\hat{u}(\mathbf{Dom}) = \sum_k^K(c_kf_k(\mathbf{Dom})),
\label{eq:bliss}
\end{equation}
where $c_i$ is the $i$-th coefficient, and $\textbf{FG}$ is the fixed grid with dimensions $128 \times 2$ (partitioned range from 0 to 1 to 128 equal intervals). All other symbols are the same as for OFormer. 

In summary, this stage preserves architectural continuity with the first stage, introduces encoder modularity for controllable efficiency-accuracy trade-offs, and resolves the cross-attention bottleneck by predicting a fixed number of coefficients, thereby delivering scalable inference without sacrificing accuracy.

It is worth noting that the encoder ($\text{Enc}$) in BLISSNet is modular and can be replaced to meet different speed-accuracy needs. Specifically, when faster training or inference is required, the encoder can be substituted with a lighter transformer (e.g., fewer blocks) or a CNN-based backbone. However, these options usually result in lower accuracy. This flexibility also allows for fair comparisons with other methods: for instance, when benchmarking against Senseiver, we can integrate its encoder into BLISSNet while maintaining the overall architecture described above. This plug-and-play capability enables systematic exploration of efficiency-performance trade-offs within a common framework.

To train the second-stage model, the total loss is defined as follows:
\begin{equation}
    \mathcal{L} = \lambda_1\mathcal{L}_{cp} + \lambda_2\mathcal{L}_{coef} + \lambda_3\mathcal{L}_{emb} + \lambda_4\mathcal{L}_{gt},
\end{equation}
where $\mathcal{L}_{cp}$ measures the discrepancy between the generated output and the known control points in the initial image, evaluated only at locations where ground-truth control points are available. This term is computed cumulatively in the same manner as the first-stage loss. $\mathcal{L}_{coef}$ is an MSE loss applied to the coefficients predicted in the first and second stages, following a student-teacher formulation. $\mathcal{L}_{emb}$ denotes the loss between the embeddings immediately preceding the coefficient decoder in both models. $\mathcal{L}_{gt}$ is the normalized MSE between the predicted output and the complete ground truth field. All weighting factors $\lambda_i$ coefficients are selected independently for a given flow type.

The detailed equations for the loss are presented below: 
\begin{equation}  
    \mathcal{L}_{\text{cp}}=\frac{1}{B}\sum_{b}^{B}\frac{1}{|S|}\sum_{i\in S}\sum_{j=1}^{K}\left( u_{\text{true}}^{(b)}(x_i)-\sum_{k}^{j} c_k^{(b)}\,f_k(x_i)\right)^2
\end{equation}
\begin{equation}
    \mathcal{L}_{coef} = \frac{1}{B}\sum_{b}^{B}\frac{||\hat{\textbf{c}}^{(b)} - \textbf{c}^{(b)}||_2^2}{K},  
\end{equation}
\begin{equation}
    \mathcal{L}_{\text{emb}} = \frac{1}{B}\sum_{b}^{B}\frac{||\widehat{\textbf{emb}}^{(b)} - \textbf{emb}^{(b)}||_2^2}{E}, 
\end{equation}
\begin{equation}
    \mathcal{L}_{gt} = \frac{1}{B}\sum_{b}^{B}\frac{||u_{\text{true}}^{(b)}(\mathbf{Dom}) - \sum_{k}^{K} c_k^{(b)}\,f_k(\mathbf{Dom})||_2}{||u_{\text{true}}^{(b)}(\mathbf{Dom})||_2} ,
\end{equation}
where $B$ is the number of samples in dataset, $K$ is the number of basis functions, $S \subseteq \{1,\dots,D^2\}$ is the set of indices of the control points, $x_i$ is the coordinate associated with index $i$, $c_k^{(b)}$ is the $k$-th coefficient of the $b$-th sample, and $\hat{\textbf{c}}^{(b)}$ and ${\textbf{c}}^{(b)}$ are vectors of the coefficients for sample $b$ for the first stage and second stage respectively, $f_k$ is the $k$-th basis function, $u_{\text{true}}$ is the target function, $\mathrm{\textbf{emb}}$ is the model embedding (output from $\text{CA}_{\text{bliss}}$ see Equation~\eqref{eq:decoder_bliss}), and $\widehat{\mathrm{\textbf{emb}}} \in \mathbb{R}^{E}$ is the stage-one embedding we want to learn from, $E$ is the embedding dimension.

\subsection{Interpolation}
To recover the full domain from sparse sensor measurements, BLISSNet performs interpolation to fill the missing regions. The input consists of sensor values along with their corresponding coordinates within the domain. The branch network processes the vector of sensor values and coordinates, while the trunk network receives a meshgrid representing the desired domain dimensions. The model then generates the reconstructed field over the target domain (Fig.~\ref{fig:BLISSNet_architechture}).

An important property of an interpolation operator such as BLISSNet is its ability to generalize across varying numbers of sensor measurements and different domain resolutions without architectural changes. The model flexibly adapts to the spatial distribution of input observations and the resolution of the target grid, producing reconstructions that remain consistent with the provided sensor data while matching the dimensions of the given meshgrid.

\subsection{Data assimilation}
Data assimilation is the process of combining observational data with model predictions to obtain an improved estimate of the system state. This estimate is providing an optimal balance between measurement information and physical consistency. Among the most widely used data assimilation approaches are Kalman filter-based methods. Although highly accurate, these techniques are computationally demanding~\cite{nudging_fast, nudging_fast_2}, as they require the inversion of large covariance matrices, which can also introduce numerical instability in high-dimensional systems.

AOT-nudging offers a computationally efficient alternative~\cite{nudging_aot}. In standard nudging, corrections are applied only at the observation points. In contrast, AOT-nudging first constructs a spatial interpolation of the observations over the domain and then applies corrections through this interpolated field, allowing sparse measurements to influence the surrounding region~\cite{aotnuging_vs_nudging}. As a result, the accuracy of AOT-nudging depends directly on the quality of the interpolation operator, for which BLISSNet provides a suitable and scalable solution.



The AOT-nudging is based on the following formula: 
\begin{equation}
\boldsymbol{x}_t = \mathcal{M}(\boldsymbol{x}_{t-1}) + \kappa [ I_h(\boldsymbol{y}_{t-1}) -  I_h(\boldsymbol{x}_{t-1})], \label{eq:nudging}
\end{equation}
where $\mathcal{M}$ represents the model that evolve the state of the system, $\kappa$ is the relaxation parameter that adjusts the spatial scales of $\boldsymbol{x}_{t-1}$ toward those of the observed data, $I_h$ is the data interpretation operator and the subscript $h$ indicates the spatial grid spacing of the data, $\boldsymbol{y}_{t-1}$ is the observed data.

Nudging is among the fastest of several common DA methods because it avoids costly matrix inversions and multiplications. Its runtime is governed by the forward model and by the interpolation operator, which is applied twice at each assimilation time when new observations are available. Therefore, we design the interpolation operator BLISSNet to serve as $I_h$ to be both fast and accurate so that nudging remains accurate while meeting real-time constraints.

\subsection{Experiments}
To compare the performance of BLISSNet with the OFormer model, we evaluate both methods on fluid flow datasets under two experimental settings. The first is sparse-sensor reconstruction, where the full field is recovered from limited measurements. The second is AOT-nudging data assimilation, where the interpolation operator is integrated into the assimilation loop. These experiments assess both standalone reconstruction accuracy and performance within a physics-coupled assimilation framework.

\subsubsection{2D Navier-Stokes}
The turbulent data for the training and testing of both interpolation models is
obtained from a numerical simulation of Navier-Stokes (NS) flow, which uses the following equations:
\begin{equation}
\begin{aligned}
    &\partial_t \psi(x,t) + v(x,t)\cdot\nabla \psi(x,t) = \nu \Delta \psi(x,t) + \phi(x), \\
    &\nabla \cdot v(x,t) = 0, \\
    &\psi(x,0) = \psi_0(x), \qquad x \in (0,1)^2, \; t \in (0,T]
\end{aligned}
\label{eq:ns}
\end{equation}
with periodic boundary condition where $v$ is the velocity field, $\psi$ is the vorticity, $\psi_0$ is inital voricity, $\nu$ is the viscosity coefficient (taken here as random from $1e-5$ to $1e-3$ for different initial conditions) and $\phi$ is a forcing function. The forcing is kept fixed: $\phi(x) = 0.1 \left( \sin\left(2\pi(x_1 + x_2)\right) + \cos\left(2\pi(x_1 + x_2)\right) \right)$, where $x=(x_1, x_2)$ represents a point on the discretized grid. The governing equations were solved using a pseudospectral method with randomly generated initial conditions. The resulting stream function $\psi$ was then discretized on a $64\times64$ grid and used to train the models~\cite{fno}.

To train the interpolation models for the NS equations, we use sparse sensors randomly distributed across the domain. For each training sample, the number of sensors is drawn uniformly between 30 and 150, and their locations are selected at random. The model receives only the flow values at these selected points, together with their spatial coordinates, and should reconstruct the full flow field. We evaluate each algorithm based on its ability to recover the complete NS state from these limited and randomly placed measurements.

\subsubsection{Quasi-Geostrophic}
We next evaluate a two-layer Quasi-Geostrophic (QG) flow to test generalization to a different flow regime:
\begin{equation}
\frac{\partial q_1}{\partial t} + \bar{v}_1 \frac{\partial q_1}{\partial x} + \frac{\partial \bar{q}_1}{\partial y} \frac{\partial \psi_1}{\partial x} + J(\psi_1, q_1) = \text{ssd}, \label{eq:qg_layer1}
\end{equation}

\begin{equation}
\frac{\partial q_2}{\partial t} + \bar{v}_2 \frac{\partial q_2}{\partial x} + \frac{\partial \bar{q}_2}{\partial y} \frac{\partial \psi_2}{\partial x} + J(\psi_2, q_2) = -R_2 \nabla^2 \psi_2 + \text{ssd}, \label{eq:qg_layer2}
\end{equation}
with periodic boundary condition, where $\text{ssd}$ refers to small-scale dissipation, which consists of higher-order derivative terms and is neglected in this context. The streamfunction in layer $i$ is $\psi_i$, the potential vorticity is $q_i$, the Jacobian is $J(a,b)$, the layer mean velocity is $\bar{v}_i$, and $R_2$ is a decay rate. We solved the governing equations using a pseudospectral method with randomly generated initial conditions, and the resulting stream function $\psi$, discretized on a $64\times64$ grid, was used for model training~\cite{qg}.

For the experiments with the QG model, sensors are also placed randomly across the domain. For each sample, the number of sensors is drawn uniformly between 150 and 250, which is higher than in the previous experiment to account for the increased complexity and finer-scale structures of the QG system. The model receives the corresponding observations and is required to reconstruct the top-layer QG field. This setup evaluates whether the interpolation method can accurately recover complex flow structures from sparse and irregular measurements.

\subsection{Training}
We generate training data by solving the governing equations: Equation~\eqref{eq:ns} for the NS and Equations~\eqref{eq:qg_layer1} --~\eqref{eq:qg_layer2} for the QG flow. For each sample, we draw a random initial condition and evolve the system forward in time. 
To avoid generating training data with strong similarity to the initial condition,
we discard the first 300 frames. 
The datasets consist of image snapshots of the stream function, with a non-dimensional temporal resolution of $\Delta t = 1e-3$ for NS and a dimensional time step of 540 seconds for QG.
The NS dataset contains (21,200/2,000/5,000) images for training, validation, and testing, respectively, while the QG dataset contains (56,919/7,520/10,000) images, corresponding to an approximate 75\%/10\%/15\% split. 

We perform a random search over major hyperparameters and select the following configuration: encoder depth $=8$; post-cross-attention feature map size $=128\times 512$, fixed grid size ($\textbf{FG}$) $=128 \times 2$, SIREN layer width $=512$, and number of basis functions $=K$ set to 512 for NS and 1024 for QG. 
For the NS experiment, the loss weights are ($\lambda_{cp}$, $\lambda_{coef}, \lambda_{emb}, \lambda_{gt})=(10, 40, 0.01, 0.05)$. 
For the QG experiment, the weights are (2, 35, 0.01, 0.01) respectively. 
We choose these values to place all the loss components on comparable amplitude scales.

We train with the Adam optimizer on randomly sampled mini-batches of size 20.
The initial learning rate is $2\times 10^{-5}$, and a ReduceLROnPlateau scheduler halves the rate after five epochs without validation-loss improvement. 
Over 120 epochs, the learning rate decreases from $2\times 10^{-5}$ to $7.81\times 10^{-8}$. All experiments were conducted on a single NVIDIA GeForce RTX 3080 Ti GPU.

\section{Results}

\subsection{Time complexity analysis}
BLISSNet scales better than cross-attention models as the domain size increases, because its computational complexity has a smaller constant factor. In OFormer, let $D\times D$ be the size of the domain grid, $F$ the feature width in cross-attention, and $G$ the output MLP width. The dominant costs are $\mathcal{O}(D^{2}F^{2}h)$ from projecting the matrix of size $(D^{2}\times F)$ by $(F\times F)$ $h$ times (where $h$ is number of heads of attention) and $\mathcal{O}(D^{2}FGh)$ from the output MLP, giving a total of $\mathcal{O}\left(D^{2}(F^{2}h+FGh)\right)$. In BLISSNet, the branch outputs a fixed-size coefficient vector independent of $D$, while the trunk with MLP width $G_{\text{trunk}}$, evaluates $K$ learned bases functions over the grid with cost $\mathcal{O}\left(D^{2} K G_{trunk}\right)$ plus basis function summation cost $\mathcal{O}\left(D^{2} K\right)$, yielding to $\mathcal{O}\left(D^{2} (K G_{trunk}+K)\right)$. The complexity can be reduced to $\mathcal{O}\left(D^{2} K\right)$ if the trunk is precomputed. With $F=512$ and $h=8$ for OFormer and $G_{trunk}=512$, $K \le 1024$ by design for BLISSNet, we have $F^{2}h+FGh > K G_{trunk}+K \gg K$. Therefore, although both methods scale as $\mathcal{O}(D^{2})$ in domain size, BLISSNet has a significantly smaller leading constant and delivers faster inference, especially when using a precomputed trunk. 



\begin{figure}[h]
    \centering
    \begin{subfigure}[b]{0.5\textwidth}
        \centering
        \includegraphics[width=\linewidth]{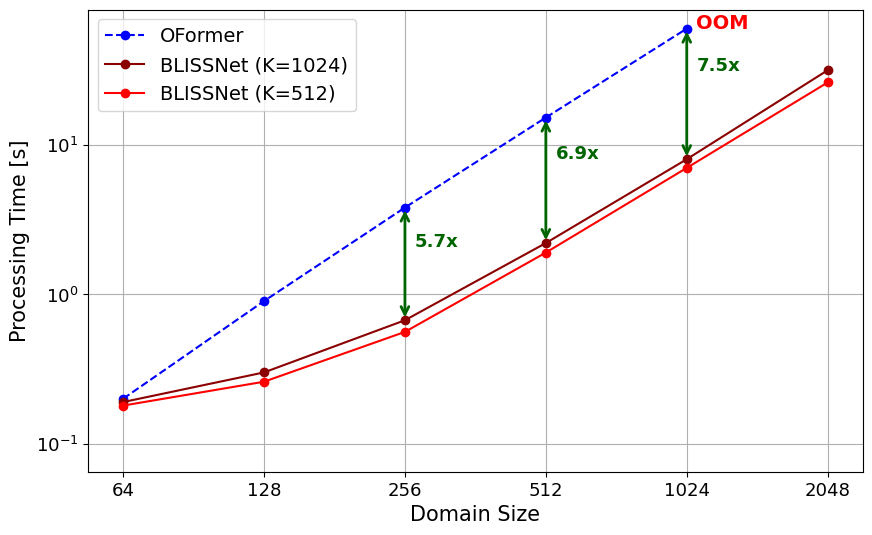}
        \caption{}
        \label{fig:NS_timecomparison_a}
    \end{subfigure}\hfill
    \begin{subfigure}[b]{0.451\textwidth} 
        \centering
        \includegraphics[width=\linewidth]{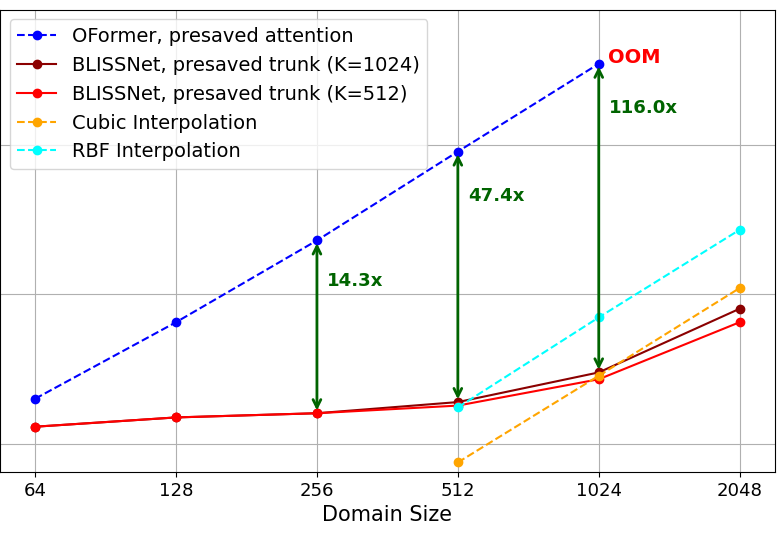}
        \caption{}
        \label{fig:NS_timecomparison_b}
    \end{subfigure}
    \caption{Time comparison of BLISSNet and OFormer. BLISSNet is more than 8 times faster on domains of size 512 and larger (a), and with precomputed bases it is up to 30 times faster (b). These results demonstrate a clear scalability advantage of BLISSNet for large domains.}
    \label{fig:NS_timecomparison}
\end{figure}

This theoretical advantage is reflected in empirical runtime measurements. As shown in Fig.~\ref{fig:NS_timecomparison_a}, BLISSNet achieves a sevenfold speedup over OFormer on domains of  512×512 and larger, with runtime measured as wall-clock time on an AMD Ryzen 9 5900X CPU. The performance gap widens further when grid-dependent components are precomputed. When the spatial grid is known in advance, matrices of projected coordinates that later fuse with sensor features can be cached. For example, in OFormer, the coordinate projection via the attention query matrix $Q$ and the added positional encodings can be computed once and reused. While in BLISSNet, the trunk that evaluates basis functions is independent of sensor values and can be fully precomputed and then combined with the coefficients from the branch network. This caching accelerates both models, but BLISSNet benefits more, achieving speedups of approximately 47–116 times on domains of 512 and larger. (Fig.~\ref{fig:NS_timecomparison_b}).

Moreover, on large domains (sizes of 1024 and above), the BLISSNet model with presaved trunk (precomputed basis function generation) outperforms bicubic interpolation implemented via \textit{scipy.interpolate.griddata} in terms of computational time. The runtime of BLISSNet could be further reduced by executing the part of the model (branch) on a GPU, and the model can also be inferred in batches when required.

An additional advantage of BLISSNet is its much lower memory footprint on large domains compared to the cross-attention OFormer. Because cross-attention operates on a constant-size vector and the trunk’s basis functions can be precomputed, BLISSNet holds fewer active tensors and performs fewer operations than domain-wide cross-attention. In our setup with a domain size $2048 \times 2048$, baseline OFormer exceeded 100 GB of RAM and failed with an out-of-memory (OOM) error, while BLISSNet completed with about 40 GB. Therefore, across various domain sizes, BLISSNet is both faster and more memory-efficient than the cross-attention OFormer approach.

\subsection{Interpolation}
\subsubsection{2D Navier-Stokes}
\begin{figure}[h]
    \centering 
    \includegraphics[width=0.9\textwidth]{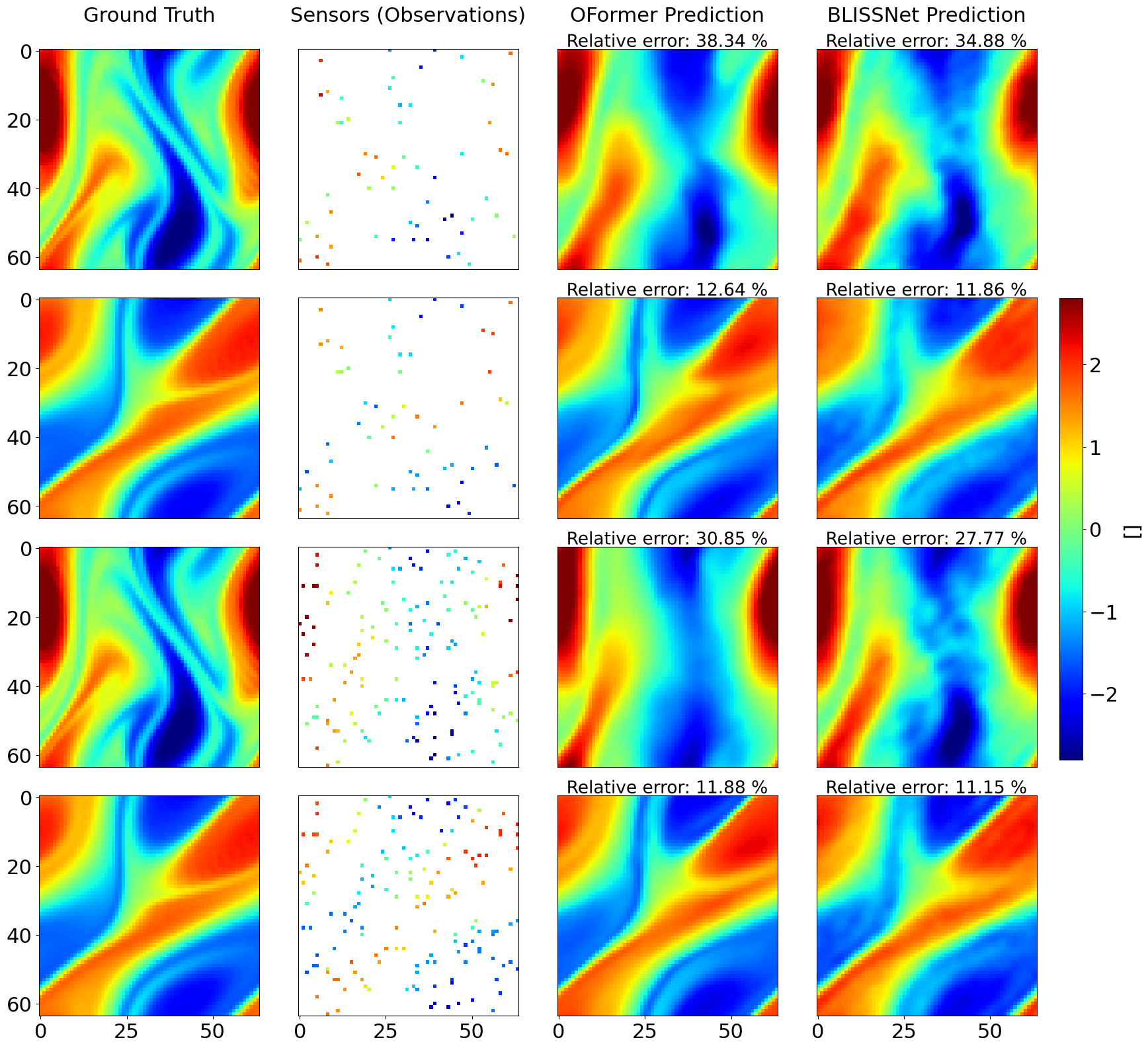}
    \caption{Comparison of interpolation performance between the baseline OFormer and our BLISSNet model using 60 (top two rows) and 150 (bottom two rows) sensors with 10\% Gaussian noise on the sensor values (first and third row are the same image, as well as second and fourth)}
    \label{fig:NS_60_150_sensors_noise}
\end{figure}

\begin{figure}[h]
    \centering 
    \includegraphics[width=0.7\textwidth]{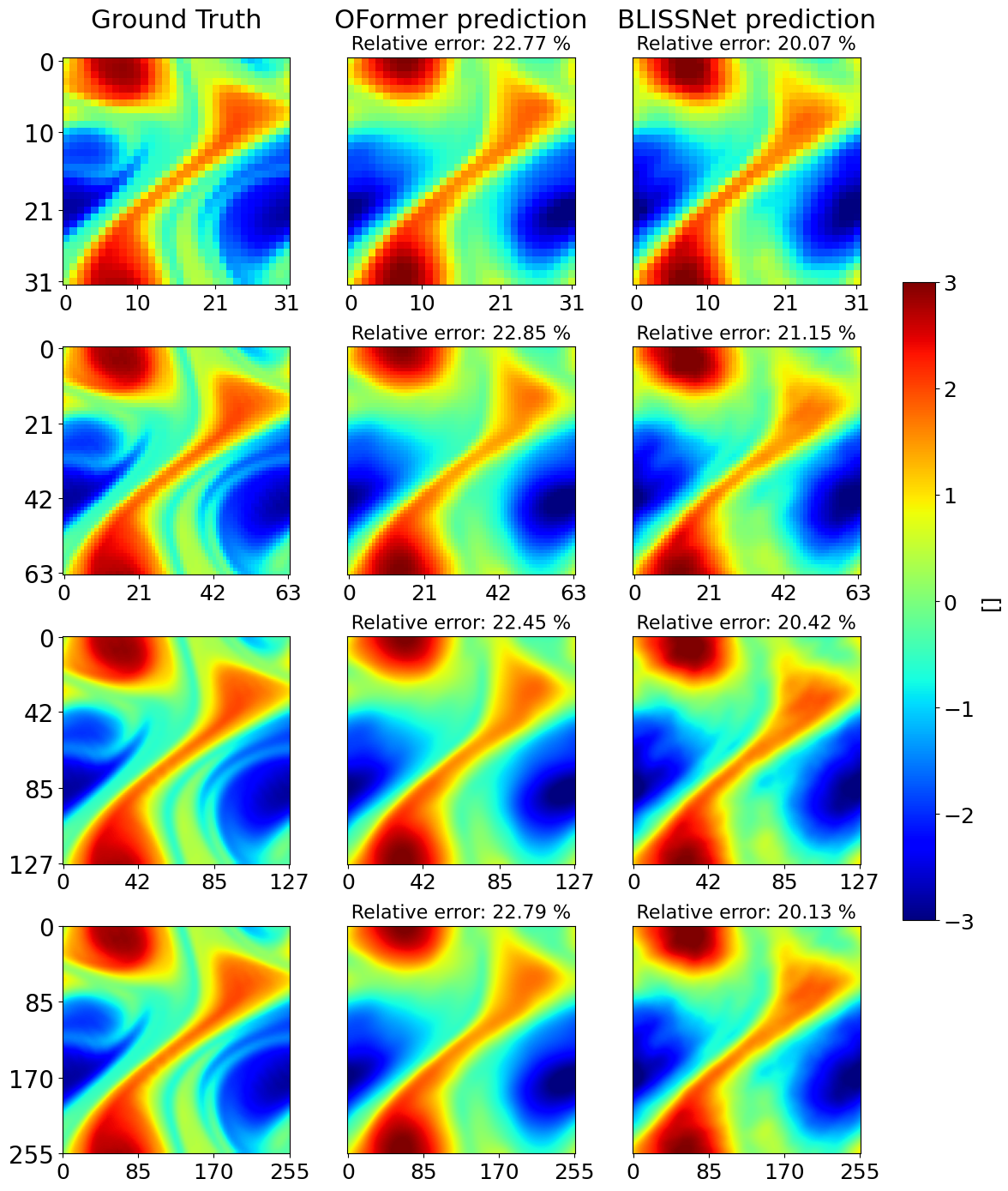}
    \caption{Comparison of interpolation performance between the baseline OFormer and our BLISSNet model using 60 sensors with 10\% Gaussian noise on the sensor values for zero-shot domain sizes, first row - $32 \times 32$ size, second row - $64 \times 64$ size, third row - $128 \times 128$ size, fourth row - $256 \times 256$ size }
    \label{fig:NS_differentratio}
\end{figure}

\begin{figure}[h!]
    \centering
    \begin{subfigure}[b]{0.52\textwidth}
        \centering
        \includegraphics[width=\linewidth]{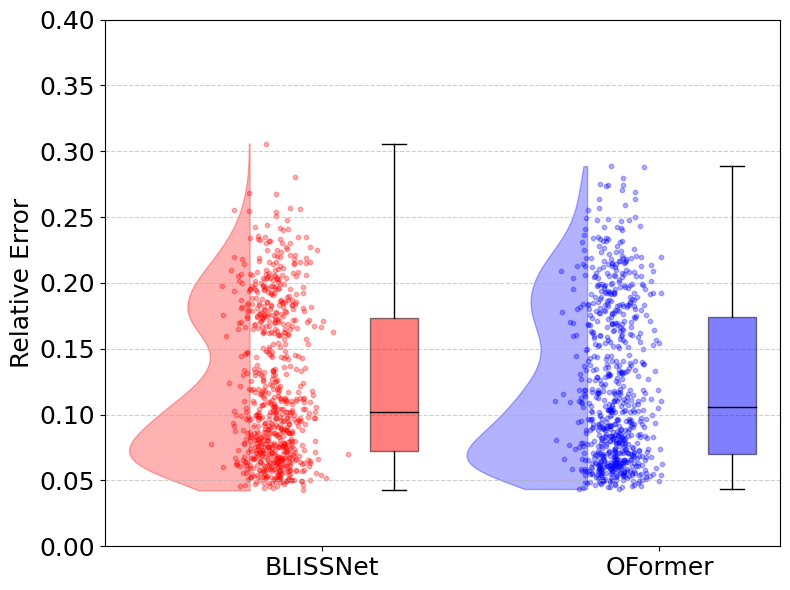}
        \caption{}
        \label{fig:NS_raincloud_60sensors}
    \end{subfigure}
    \hfill
    \begin{subfigure}[b]{0.452\textwidth}
        \centering
        \includegraphics[width=\linewidth]{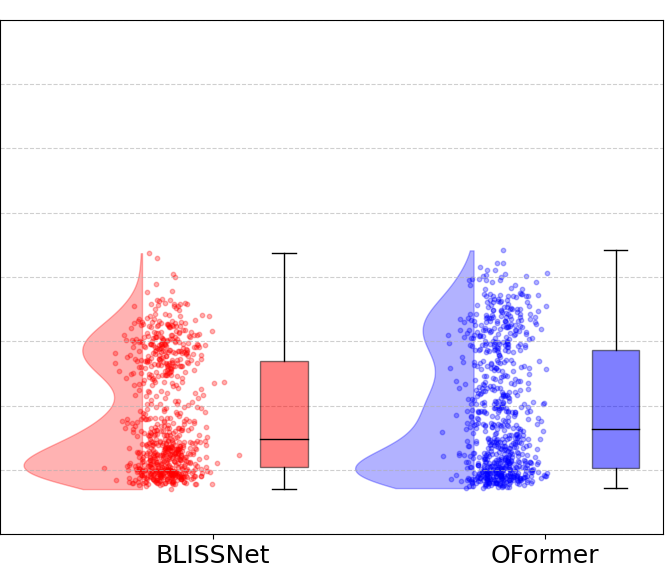}
        \caption{}
        \label{fig:NS_raincloud_150sensors}
    \end{subfigure}
    \hfill
    \begin{subfigure}[b]{0.52\textwidth}
        \centering
        \includegraphics[width=\linewidth]{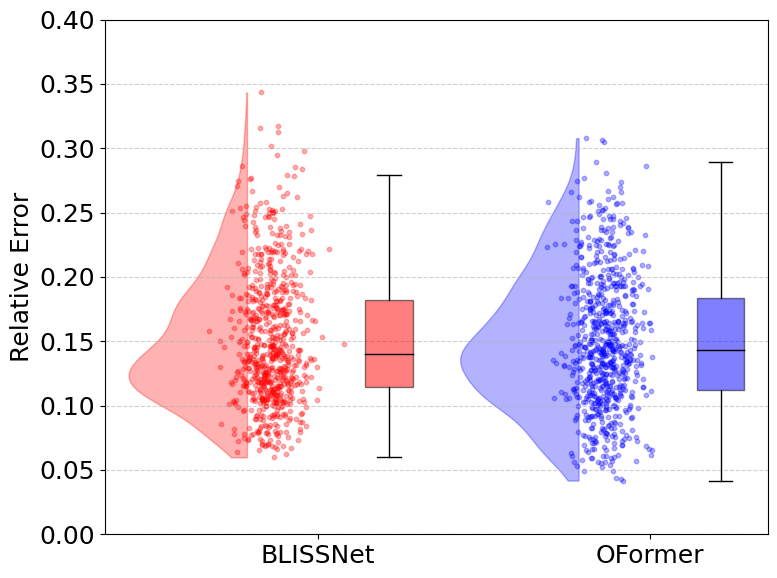}
        \caption{}
        \label{fig:NS_raincloud_60sensors_128}
    \end{subfigure}
    \hfill
    \begin{subfigure}[b]{0.452\textwidth}
        \centering
        \includegraphics[width=\linewidth]{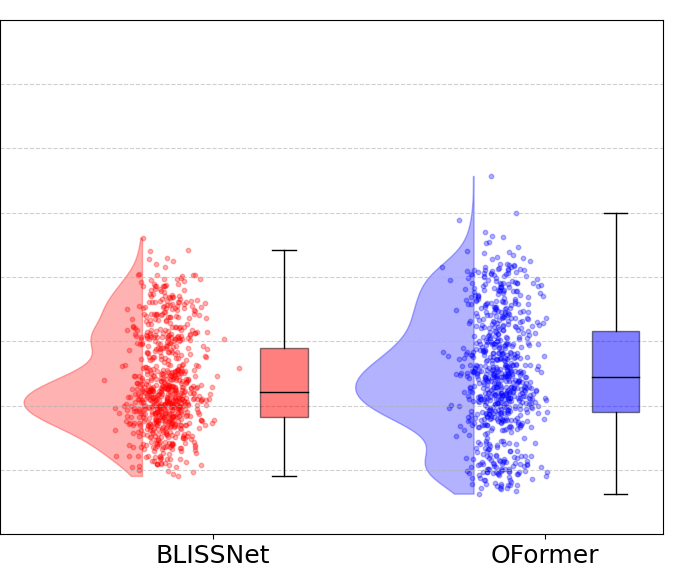}
        \caption{}
        \label{fig:NS_raincloud_150sensors_128}
    \end{subfigure}

    \caption{Boxplot comparison between OFormer and BLISSNet across different sensor configurations and domain sizes for NS flow. (a) - 60 sensors for $64\times64$ domain, (b) - 150 sensors for $64\times64$, (c) - 60 sensors for $128\times128$ domain, (d) - 150 sensors for $128\times128$ domain}
    \label{fig:NS_raincloud_all}
\end{figure}

We evaluate NS reconstruction at two sensor budgets, 60 and 150 sensors, and found that BLISSNet consistently outperforms OFormer in terms of reconstruction error. From the simulated flow trajectory, we select 30 random images and, for each image, generate 25 sensor configurations containing either 60 or 150 sensors. 
To simulate realistic measurement conditions, we add 10\% zero-mean Gaussian noise to all synthetic sensor measurements. 
Figs.~\ref{fig:NS_raincloud_60sensors} and~\ref{fig:NS_raincloud_150sensors} present the resulting error distributions as raincloud plots with accompanying boxplots. 
Representative reconstructions are shown in Fig.~\ref{fig:NS_60_150_sensors_noise}, including the ground truth, sensor locations, and both model outputs for the 60 and 150 sensor settings. 
Additional examples on unseen domain size are provided in Figs.~\ref{fig:NS_raincloud_60sensors_128}and~\ref{fig:NS_raincloud_150sensors_128}, where the same sensor budgets are applied on a grid twice as large, $128 \times 128$. 
The examples of zero-shot domain-size reconstruction are illustrated in Fig.~\ref{fig:NS_differentratio}. 
Across all sensor budgets and noise conditions, BLISSNet achieves consistently lower relative errors and exhibits fewer high-error outliers, demonstrating superior reconstruction quality and greater robustness to sensor permutations compared to the OFormer baseline.

BLISSNet reconstructions in the figures above appear less smooth than those from OFormer. 
This occurs because stage one learns with only $K$ basis functions and is constrained to optimize the cumulative loss described in Sec.~\ref{sec:blissnet}, so the best reconstruction can yield recovery of less smoothness. 
Incorporating an explicit smoothness regularizer into the stage one loss would produce smoother bases and outputs, which the stage two interpolator would subsequently inherit.

\subsubsection{Quasi-Geostrophic}

\begin{figure}[h!]
    \centering
    \begin{subfigure}[b]{0.518\textwidth}
        \centering
        \includegraphics[width=\linewidth]{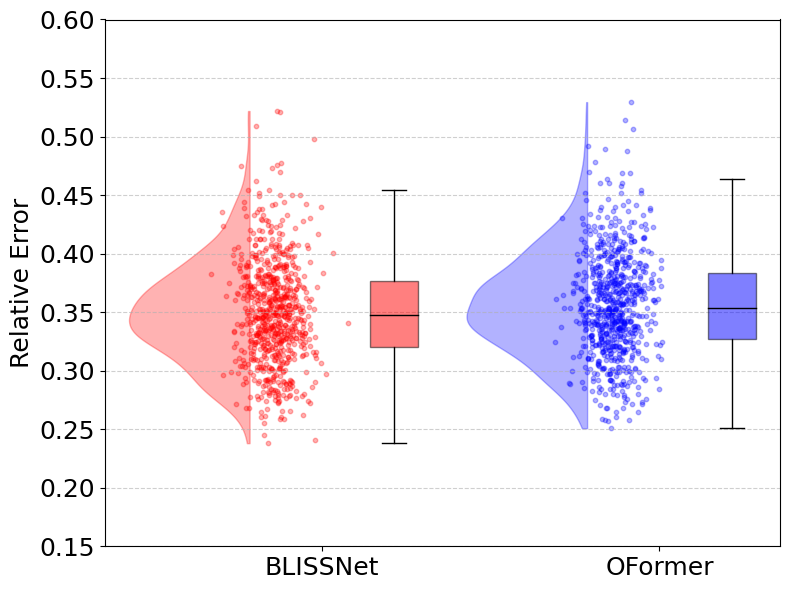}
        \caption{}
        \label{fig:QG_raincloud_150sensors}
    \end{subfigure}
    \hfill
    \begin{subfigure}[b]{0.47\textwidth}
        \centering
        \includegraphics[width=\linewidth]{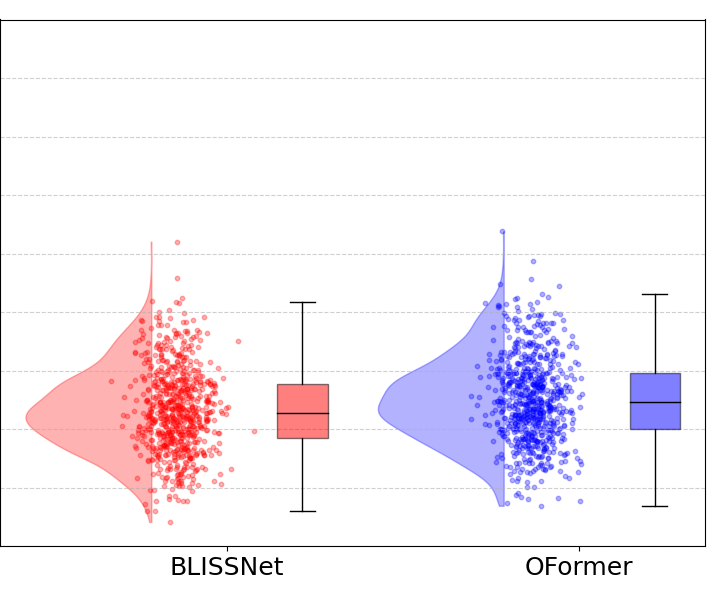}
        \caption{}
        \label{fig:QG_raincloud_200sensors}
    \end{subfigure}
    \hfill
    \begin{subfigure}[b]{0.519\textwidth}
        \centering
        \includegraphics[width=\linewidth]{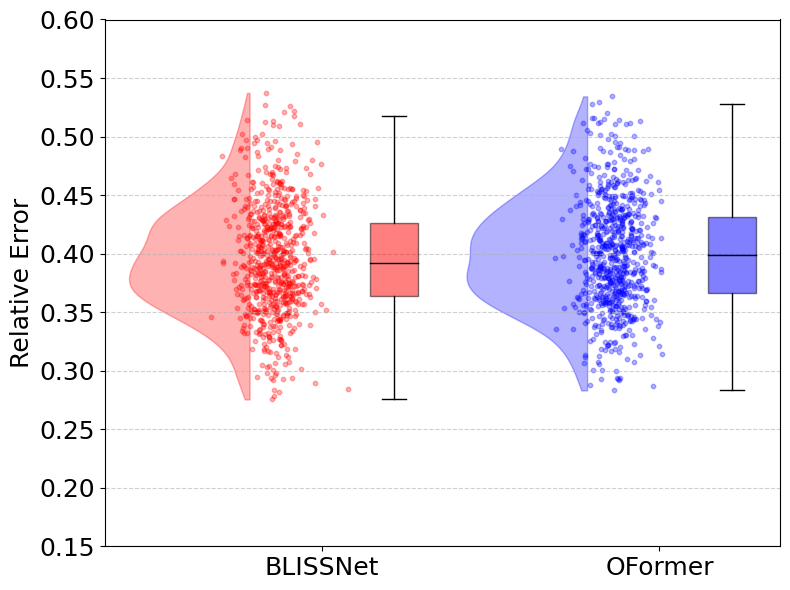}
        \caption{}
        \label{fig:QG_raincloud_150sensors_128}
    \end{subfigure}
    \hfill
    \begin{subfigure}[b]{0.47\textwidth}
        \centering
        \includegraphics[width=\linewidth]{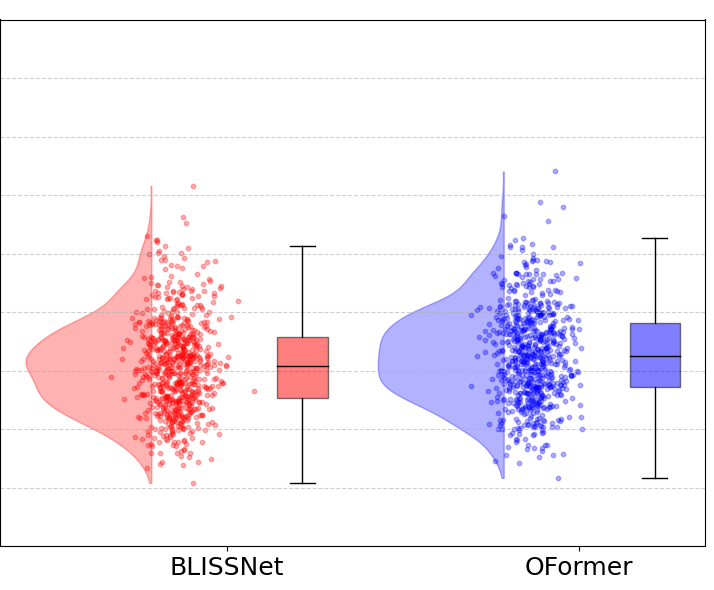}
        \caption{}
        \label{fig:QG_raincloud_200sensors_128}
    \end{subfigure}

    \caption{Boxplot comparison between OFormer and BLISSNet across different sensor configurations and domain sizes for QG flow. (a) - 150 sensors for $64\times64$ domain, (b) - 200 sensors for $64\times64$ domain, (c) - 150 sensors for $128\times128$ domain, (d) - 200 sensors for $128\times128$ domain}
    \label{fig:NS_boxplot_all}
\end{figure}

\begin{figure}[h]
    \centering 
    \includegraphics[width=0.9\textwidth]{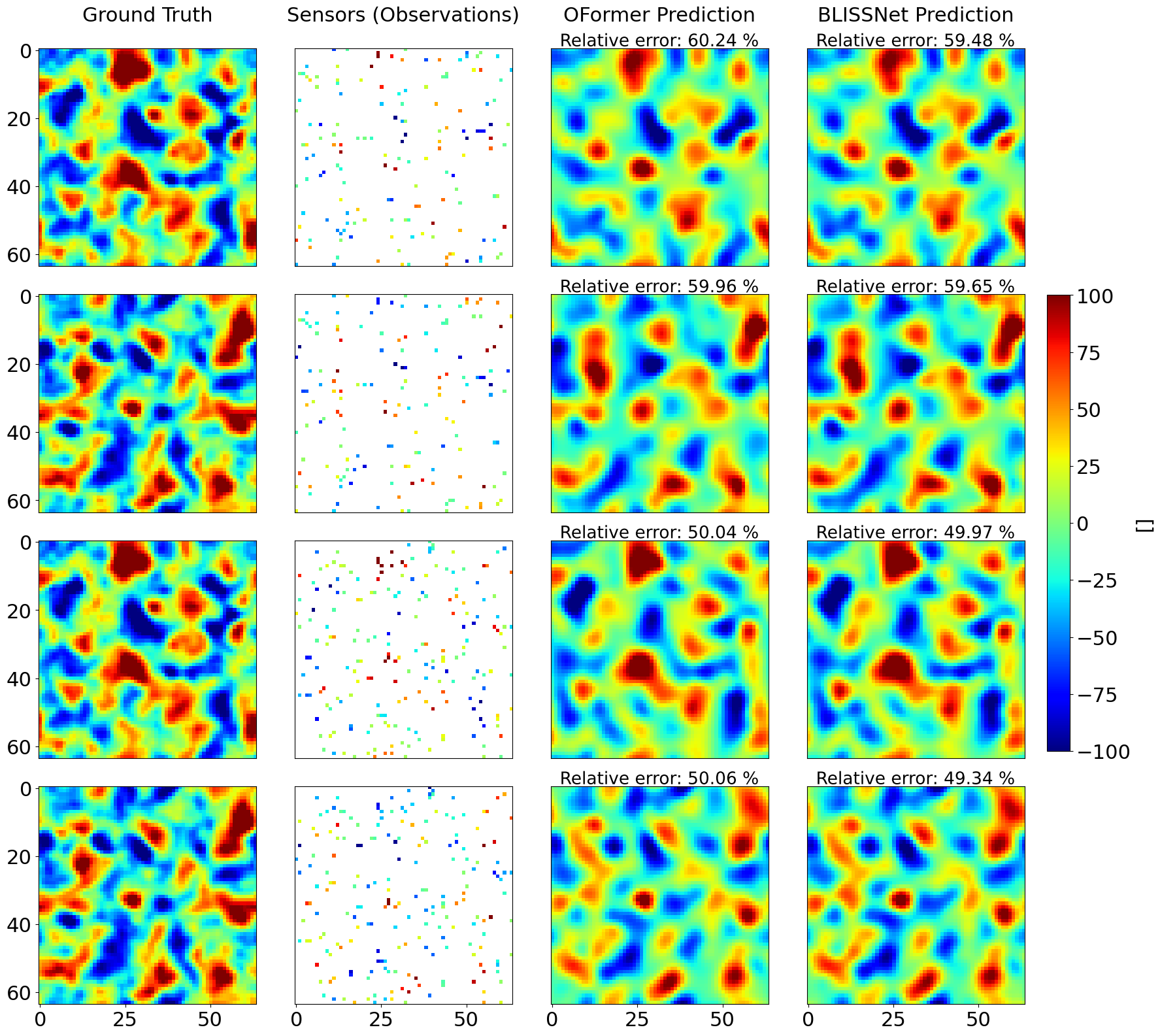}
    \caption{Comparison of interpolation performance between the baseline OFormer and our BLISSNet model using 150 (top two rows) and 200 (bottom two rows) sensors with 10\% Gaussian noise on the sensor values (first and third row are the same image, as well as second and fourth)}
    \label{fig:QG_150_200_sensors_noise}
\end{figure}

\begin{figure}[h]
    \centering 
    \includegraphics[width=0.7\textwidth]{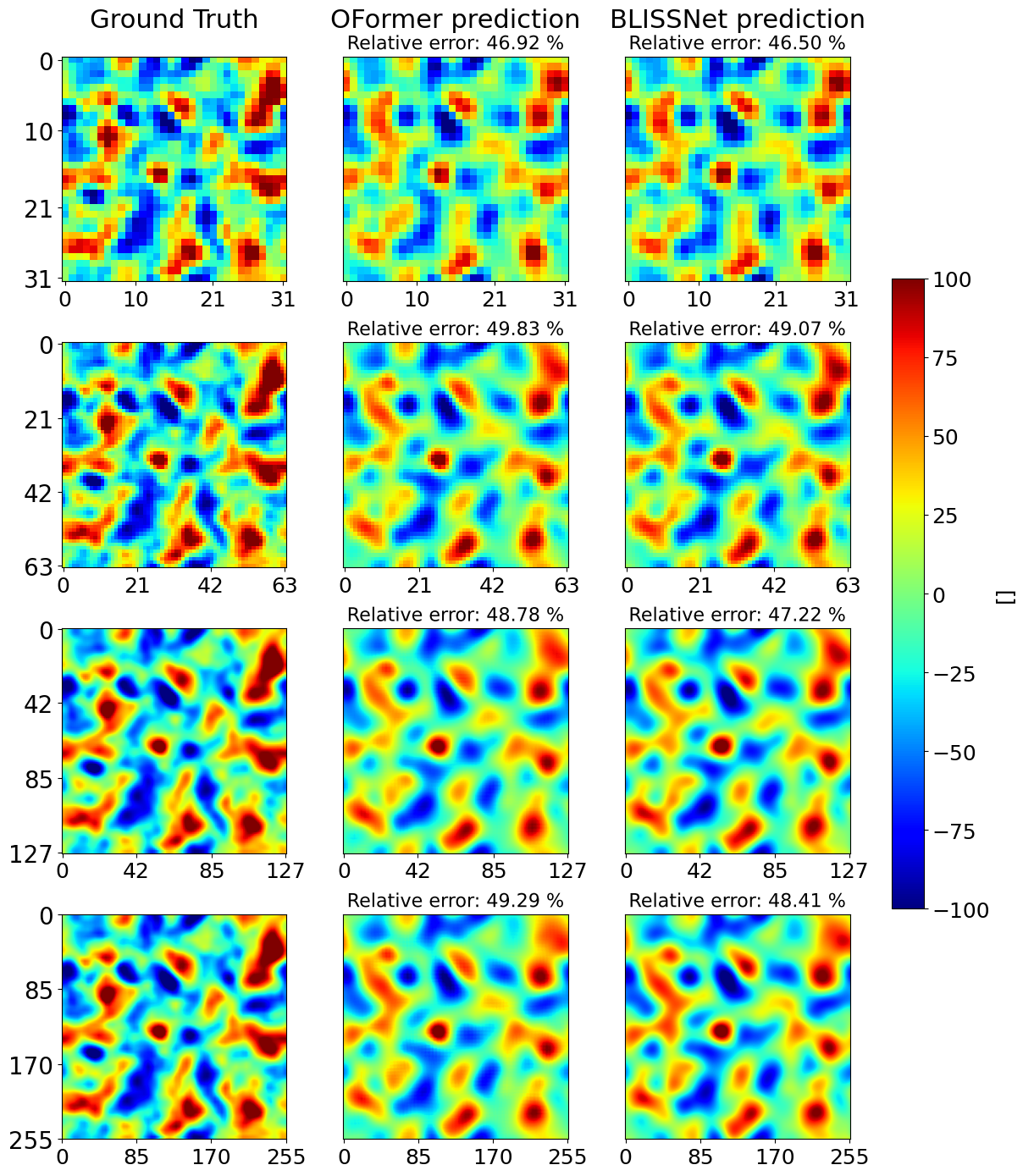}
    \caption{Comparison of interpolation performance between the baseline OFormer and our BLISSNet model using 200 sensors with 10\% Gaussian noise on the sensor values for zero-shot domain sizes: first row - $32 \times 32$, second row - $64 \times 64$, third row - $128 \times 128$, fourth row - $256 \times 256$ }
    \label{fig:QG_differentratio}
\end{figure}

The QG flow contains more complex and finer-scale features than that of the NS case, which makes it necessary to use more sensors, here 150 and 200, for reliable reconstruction.
We train and test the models to interpolate only the upper layer of the two-layer QG flow, since in realistic observational settings, only the top layer is accessible to sensors. 
In this setting, we evaluate performance by selecting 30 random snapshots and creating 25 different sensor layouts for each snapshot and sensor budget, all with 10\% Gaussian noise. 
The resulting error distributions are shown as raincloud plots with boxplots in Figs.~\ref{fig:QG_raincloud_150sensors} and \ref{fig:QG_raincloud_200sensors}, while example reconstructions, including ground truth, sensor positions, and model outputs for the 60 and 150 sensor cases, are shown in Fig.~\ref{fig:QG_150_200_sensors_noise}. Additional tests on a larger, unseen domain ($128 \times 128$) appear in Figs.~\ref{fig:QG_raincloud_150sensors_128} and \ref{fig:QG_raincloud_200sensors_128}, with zero-shot examples illustrated in Fig.~\ref{fig:QG_differentratio}. Overall, these results indicate, similar to the case with NS, that BLISSNet consistently achieves equal or lower relative errors and produces fewer high-error outliers compared to the SOTA methods, demonstrating improved reconstruction accuracy and robustness to changes in sensor placement.

\subsection{AOT-Nudging data assimilation}
\subsubsection{2D Navier-Stokes}\label{sec:ns_nudging}

\begin{figure}[h]
    \includegraphics[width=\textwidth]{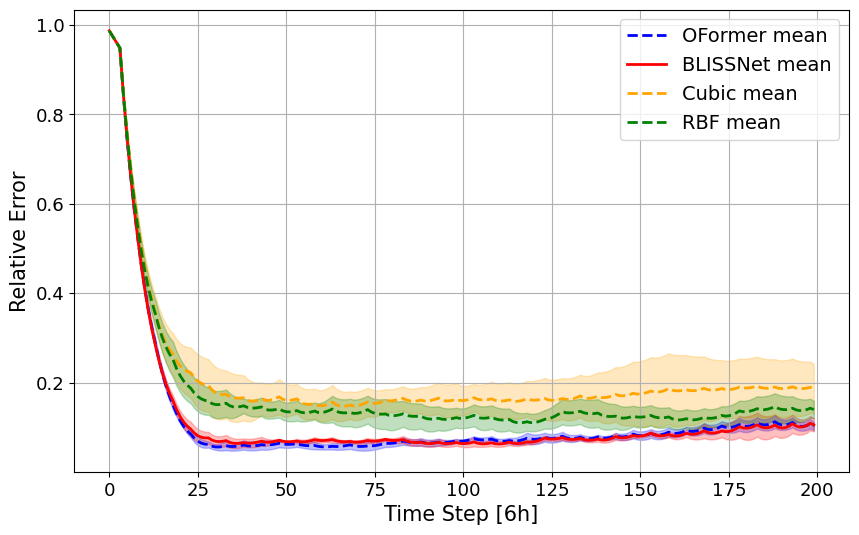}
    \caption{Comparison of interpolators in an AOT-nudging DA setup using the NS equations as the propagation model for 60 sensors. OFormer and BLISSNet show comparable relative error. Bicubic and RBF interpolations are included for reference and perform much worse than the learned interpolators. All observed data include 10\% noise.}
    \label{fig:nudging_NS_errorcomparison}
\end{figure}

\begin{figure}[h]
    \centering 
  
    \includegraphics[width=\textwidth]{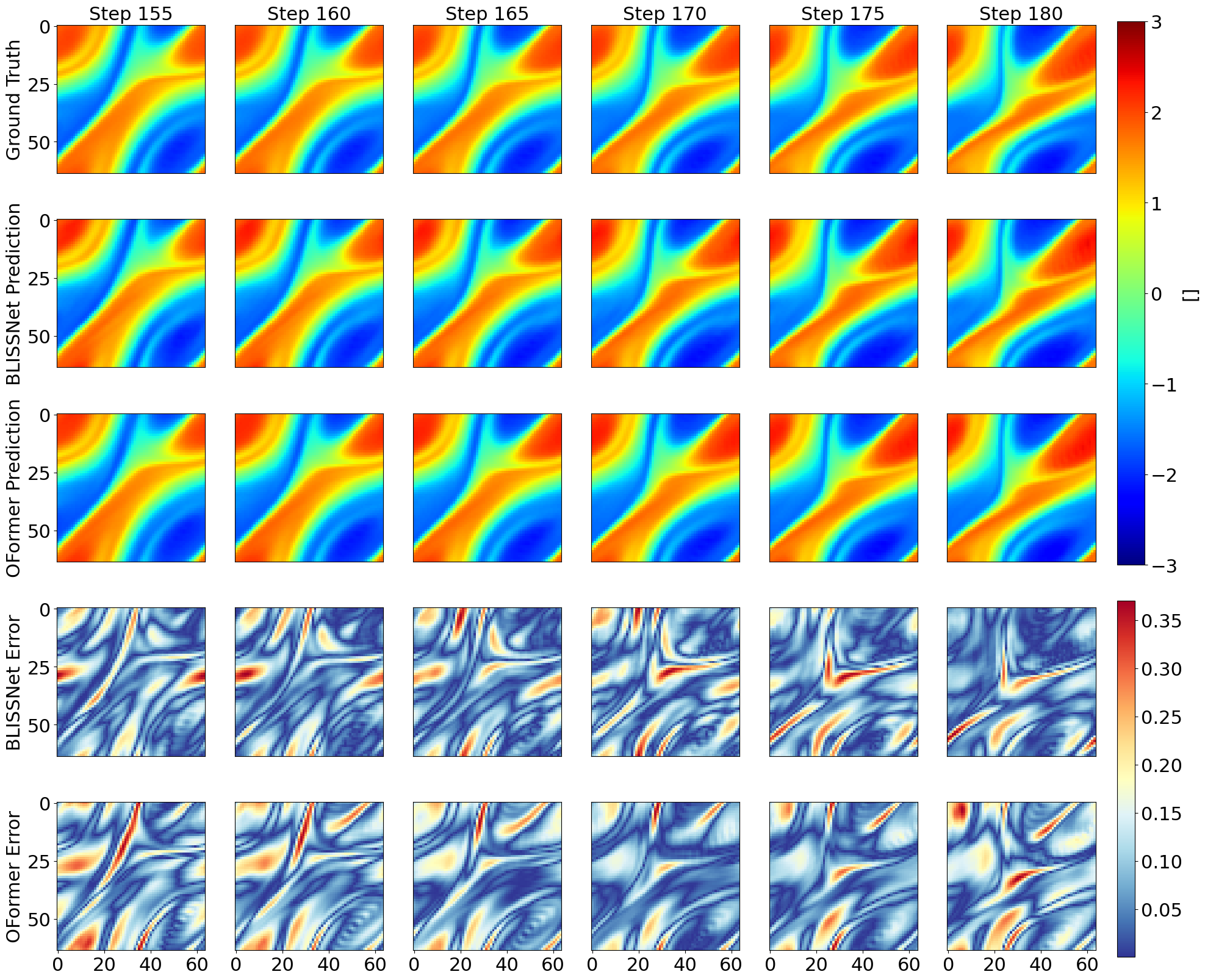}
   
    \caption{Comparison of the error of BLISSNet and OFormer for AOT-nudging}
    \label{fig:nudging_NS_errorcomparison}
\end{figure}

Within the AOT-nudging framework, which requires interpolation, employing BLISSNet-based interpolation yields superior performance compared with classical bicubic and RBF interpolation methods, while achieving accuracy comparable to that of OFormer-based interpolation.
As shown in Fig.~\ref{fig:nudging_NS_errorcomparison}, BLISSNet and OFormer yield comparable error levels, and both substantially outperform the bicubic and RBF interpolation methods when employed as interpolation operators. We sweep the nudging coefficient $\kappa$ over the range $0.01$–$0.5$ and find that $\kappa = 0.1$ provides the best results for all models.

\subsubsection{Quasi-Geostrophic}

\begin{figure}[h]
    \includegraphics[width=0.9\textwidth]{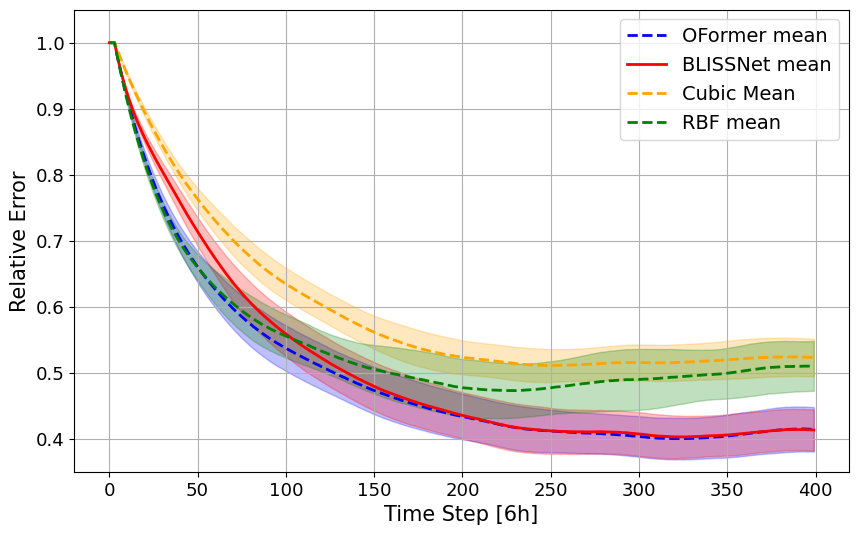}
    \caption{Comparison of interpolators in an AOT-nudging DA setup using the QG equations as the propagation model for 150 sensors. OFormer and BLISSNet show comparable relative error. Bicubic and RBF interpolations are included for reference and perform much worse than the learned interpolators. All observed data include 10\% noise.}
    \label{fig:nudging_QG_errorcomparison}
\end{figure}

We conducted an experiment analogous to Sec.~\ref{sec:ns_nudging} to compare interpolation models within the AOT-nudging framework for the QG flow configuration for 150 sensors (Fig.~\ref{fig:nudging_QG_errorcomparison}). The neural models outperformed bicubic and RBF interpolation and showed comparable performance to each other. For a fair comparison, the nudging weight coefficient $\kappa$ was tuned for each method, yielding $\kappa = 0.02$ for BLISSNet and OFormer and $\kappa = 0.05$ for bicubic and RBF.

Figure \ref{fig:nudging_QG_errorcomparison} indicates that BLISSNet and OFormer exhibit highly similar error patterns. These similarities suggest that the proposed model maintains consistent interpolation performance across datasets, achieving accuracy comparable to the OFormer baseline. Notably, the converged error level for the QG model is higher than for the NS case. This difference can be attributed to the increased complexity of the QG system, which contains finer-scale structures and two interacting layers. Since observations are available only from the top layer, the bottom layer remains unobserved and can introduce additional reconstruction error. Overall, the results support the conclusion that BLISSNet provides reliable and robust interpolation capabilities on par with established methods.

\begin{figure}[h]
    \centering 
    \includegraphics[width=\textwidth]{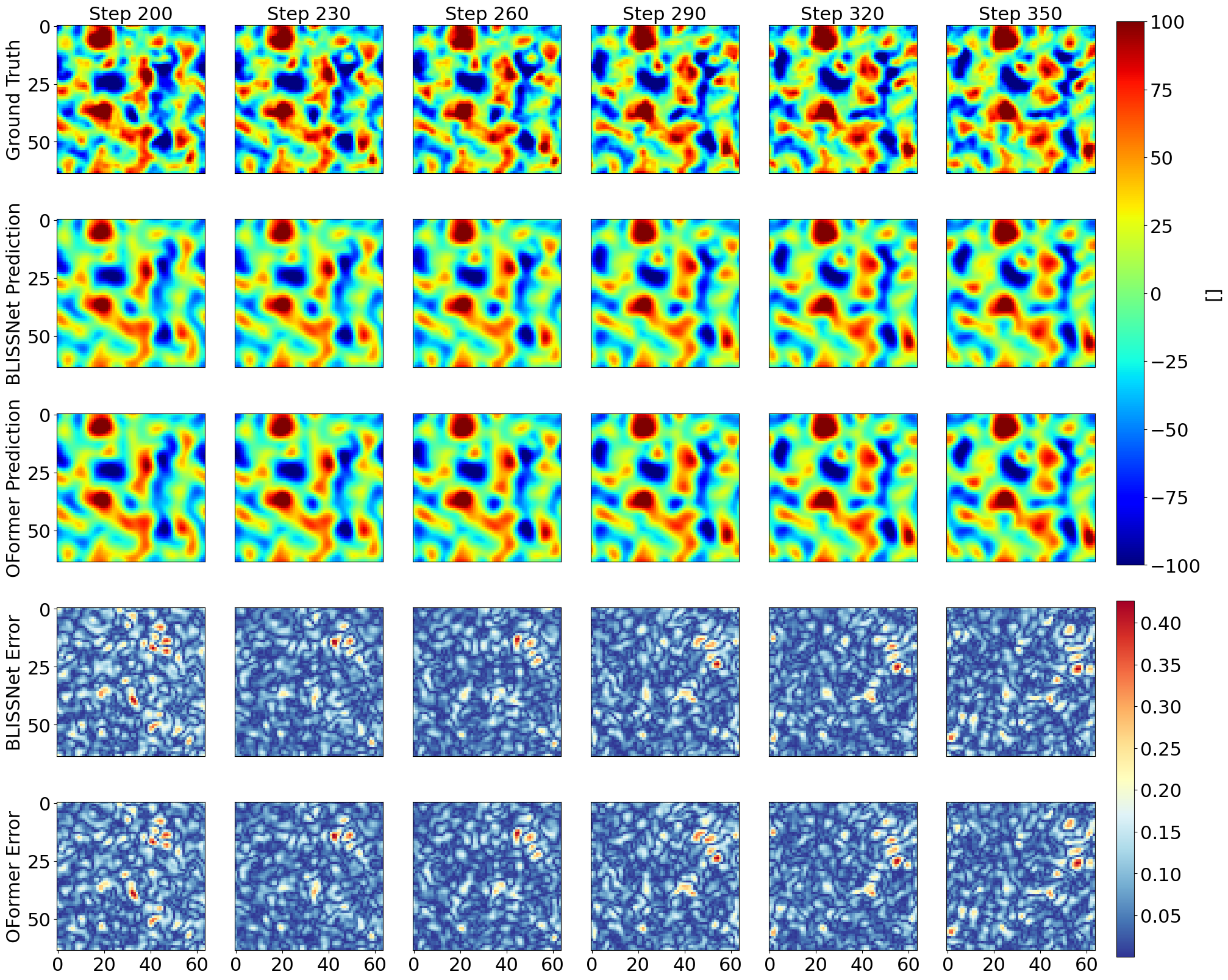}
    \caption{Comparison of the error of BLISSNet and OFormer for AOT-nudging for QG flow}
    \label{fig:nudging_NS_errorcomparison}
\end{figure}

\section{Concluding remarks}
In this paper, we introduce BLISSNet, a model for interpolation from sparse observations over a square domain. 
Our experiments demonstrate that BLISSNet achieves accuracy comparable to the SOTA OFormer model while being substantially faster, up to $7.5$ times in general, and up to $116$ times when a portion of the model can be precomputed. 
All experiments were conducted under 10\% Gaussian noise. 

BLISSNet also generalizes effectively to zero-shot domain sizes. Although trained on $64\times64$ grids, it maintains strong predictive accuracy on domains of arbitrary size, achieving effective super-resolution.
Furthermore, the method can be applied to AOT-nudging data assimilation, where it achieves performance comparable to the OFormer baseline. 

Despite these advantages, the proposed model has three primary limitations.
First, training is slower, as BLISSNet requires a two-stage training procedure that increases total training time.
Second, the second stage is constrained by the first-stage output. For example, if the first stage produces blurry predictions, the second stage tends to reproduce these artifacts. This coupling also means that adding more sensors to an already sensor-dense domain provides limited benefit—the first stage limits the achievable accuracy. The cross-attention baseline model does not suffer from this issue.
Third, the model is sensitive to the choice of loss function coefficients. Different weight initializations in the first stage may produce outputs with varying magnitudes, requiring careful tuning of the loss weights to maintain consistent gradient scales.

Overall, for systems requiring high-fidelity flow reconstruction from sparse measurements, BLISSNet offers a compelling accuracy-efficiency tradeoff and demonstrates strong potential for real-time deployment, particularly when offline training cost is not a primary constraint.
Moreover, when the computational domain is known a priori or reused across repeated runs, a substantial portion of the model can be precomputed offline. This amortization reduces the online stage to a lightweight evaluation, enabling performance that can surpass even classical interpolation methods for large-scale grids (e.g., $512 \times 512$ and above).

\printbibliography
\cleardoublepage

\end{document}